\documentclass[conference]{IEEEtran}
\IEEEoverridecommandlockouts
\usepackage{cite}
\usepackage{amsmath,amssymb,amsfonts}
\usepackage{algorithmic}
\usepackage{multirow}
\usepackage{subcaption}
\usepackage{graphicx}
\usepackage{textcomp}
\usepackage{xcolor}
\def\BibTeX{{\rm B\kern-.05em{\sc i\kern-.025em b}\kern-.08em
    T\kern-.1667em\lower.7ex\hbox{E}\kern-.125emX}}
\begin{document}

\title{COVID-19 Pandemic Prediction using Time Series Forecasting Models \\
{\footnotesize \textsuperscript{*}Note: This paper is accepted in the 11th ICCCNT 2020 conference. The final version of this paper will appear in the conference proceedings.}
}

\author{\IEEEauthorblockN{Naresh Kumar}
\IEEEauthorblockA{\textit{Department of Information Technology} \\
\textit{Delhi Technological University}\\
Delhi, India \\
naresh.mehra1987@gmail.com}
\and
\IEEEauthorblockN{Seba Susan}
\IEEEauthorblockA{\textit{Department of Information Technology} \\
\textit{Delhi Technological University}\\
Delhi, India \\
seba\_406@yahoo.in}
}

\maketitle

\begin{abstract}
Millions of people have been infected and lakhs of people have lost their lives due to the worldwide ongoing novel Coronavirus (COVID-19) pandemic. It is of utmost importance to identify the future infected cases and the virus spread rate for advance preparation in the healthcare services to avoid deaths. Accurately forecasting the spread of COVID-19 is an analytical and challenging real-world problem to the research community. Therefore, we use day level information of COVID-19 spread for cumulative cases from whole world and 10 mostly affected countries; US, Spain, Italy, France, Germany, Russia, Iran, United Kingdom, Turkey, and India. We utilize the temporal data of coronavirus spread from January 22, 2020 to May 20, 2020. We model the evolution of the COVID-19 outbreak, and perform prediction using ARIMA and Prophet time series forecasting models. Effectiveness of the models are evaluated based on the mean absolute error, root mean square error, root relative squared error, and mean absolute percentage error. Our analysis can help in understanding the trends of the disease outbreak, and provide epidemiological stage information of adopted countries. Our investigations show that ARIMA model is more effective for forecasting COVID-19 prevalence. The forecasting results have potential to assist governments to plan policies to contain the spread of the virus.
\end{abstract}


\section{Introduction}
The novel Coronavirus (COVID-19) has infected millions of people worldwide since it emerged from China in December 2019. COVID-19 has very high mutating capability, and it can spread very easily. Infected people from this virus suffer from severe respiratory problems, and may develop serious illness if suffering from chronic diseases like cardiovascular disease or diabetes or having weak immune system or being older in age \cite{sohrabi2020world}. World health organization (WHO) declared on 11th March, 2020, the outbreak of COVID-19 as a pandemic. There are challenges to contain the disease because an infected person shows symptom after a long time or no sign of the disease. At present, no vaccination has been discovered for COVID-19. In this situation, social distancing, identifying the positive cases using testing at large scale, and containment of infected person is the only option to prevent the spreading of the virus \cite{haghani2020scientific}. 

The spread of COVID-19 can be classified \cite{jia2020prediction} under three major stages- 1. Local outbreak: at this stage, spreading chain of the virus among the people can be tracked, and the source of infection can be found out. The cases in this stage mostly relate to within family or friends, or the local exposure. 2. Community transmission: at this stage, source of the chain of infected people cannot be found out. The infected cases grow through cluster transmission in the communities. 3. Large scale transmission: at this stage, the virus spreads rapidly to other regions of a country due to uncontrolled mobility of people at large scale. 

Due to high scale community impact and easy spreading worldwide, national governments imposed lockdown to control the spread of corona virus. As of 20th May, 2020, 4996472 cases have been confirmed, 1897466 cases have recovered, 2328115 deaths have been reported, and 2770891 active cases have been identified worldwide. The statistical data is collected from \cite{dong2020interactive}, and the number of COVID-19 cases is calculated between 22 Jan, 2020 to 20 May 2020. 

As no vaccine has been discovered of the disease, so motivation behind this paper is to model spreading of the corona virus, and predict the impact to optimize the planning to manage the various services and resources for the public by the governments. Some of the studies \cite{zhang2020predicting, ribeiro2020short, pal2020neural, mandal2020model} have been published showing statistical analysis, modeling, and artificial intelligence to contain the spread of the virus, and highlight impacts in coming days. These early studies are carried out using very limited information available at early stage of the outbreak. Now, the virus has spread at large scale, and much information is available for the analysis. Predictive analysis of COVID-19 has become a hot research area to support health services and governments to plan and contain the spread of the infectious disease \cite{covid2020forecasting}. Modeling and forecasting the daily spread behavior of the virus can assist the health systems to be ready to accommodate the upcoming number of patients. Accurate forecasting of the disease is a matter of concern because it may impact government’s policy, containment rules, health system, and social life. Regarding this context, we explore the predictive capability of the ARIMA \cite{wulff2017time}, and Prophet \cite{taylor2018forecasting} forecasting models. The models are widely used and accepted due to their more accurate forecasting capability. We use the day level cumulative cases of COVID-19 worldwide and 10 mostly affected countries; US, Spain, Italy, France, Germany, Russia, Iran, United Kingdom, Turkey, and India for our analysis study. 

The objective of this paper is to provide evaluative study of prediction models using COVID-19 cases, and forecasting the impact of the virus in the affected countries, and worldwide. We present trend analysis of COVID-19 cases, and compared the performance of the models using the metrics such as the mean absolute error (MAE), root mean square error (RMSE), root relative squared error (RRSE), and mean absolute percentage error (MAPE). We generate forecasting results for COVID-19 confirmed, active, recovered, and death cases. The results show that ARIMA outperformed the Prophet model. 

The rest of the paper is organized as follows. Section II presents a literature survey. Section III provides trend analysis of COVID-19 cases.  Section IV describes overview of time series forecasting models. Modeling framework and used COVID-19 dataset are described in section V. Statistical analysis and model evolution are presented in section VI. Section VII concludes the paper.

\section{Literature survey}
Intensive research work is going on to evaluate and contain the worldwide disaster of COVID-19 on the human race. Research studies include predictions about the future cases \cite{zhang2020predicting}, and analysis of the variables responsible for spread of the coronavirus \cite{oliveiros2020role}. 

In the literature, time series forecasting problems have been studied widely in which COVID-19 forecasting is an emerging problem. Forecasting models can be used to forecast the impact of the disease on the community which can help to control the epidemic. In \cite{ribeiro2020short}, authors have performed forecasting evaluation study of the models using COVID-19 day level cases from 10 mostly affected states from Brazil. According to the authors, the stacking ensemble and SVR performed better as compared to ARIMA, CUBIST, RIDGE, and RF models for the adopted criteria. In \cite{fanelli2020analysis}, the author has developed ARIMA(p,d,q) model and studied the COVID-19 epidemiological trend in the three most affected countries; Spain, Italy, and France of Europe continent using the data between 21 Feb to 15 April 2020. The author studied the various orders (p, d, q) of the model, and selected best performing order based on lowest values of MAPE for the three countries. He has suggested that ARIMA models are suitable for forecasting the COVID-19 prevalence for the upcoming days. Chintalapudi \textit{et al.} \cite{chintalapudi2020covid}, adopted seasonal ARIMA model for forecasting of COVID-19 cases in Italy using the data till 31st March 2020. They have analysed the impact of two months lockdown in Italy, and observed decrement in the confirmed cases and growth in the recovered cases due to lockdown. Alabi \textit{et al.} \cite{alabi2020covid} have adopted the Facebook Prophet model to forecast spread of COVID-19. They have performed prediction for confirmed and death cases. Their forecasting accuracy of Prophet was 79.6\% for the data from WHO between 7th April to 3rd May 2020. Parikshit \textit{et al.} \cite{mahalle2020data} have presented medical perspective of COVID-19, and prediction using Prophet model. They have recommended Prophet for prediction due to open source algorithm, accuracy, and faster data fitting. Using the Prophet model, they have predicted infected cases worldwide as 1.6 million by the end of May 2020,  and 2.3 million by the end of June 2020.

\section{COVID-19 Trends}
We have collected cumulative day level data of COVID-19 cases from github repository \cite{jhu_usa} latest by 20 May, 2020. The repository is supported by ESRI living atlas team, Applied Physics Lab (APL), and maintained by the Center for Systems Science and Engineering (CSSE), both at Johns Hopkins University, USA. The repository contains worldwide COVID-19 reported cases starting from 22 January, 2020 on day-to-day basis. We study COVID-19 confirmed cases, recovered cases, death cases, and active cases for 10 adopted countries and worldwide. The adopted countries are the badly affected countries by the virus in the world latest by 20 May, 2020.

The impact of coronavirus in 10 adopted countries from 1st March, 2020 to 20th May, 2020 is shown in Fig.  1. Trends of confirmed, recovered, deaths, and active cases in adopted countries show that impact of the virus from highest to lowest is in the sequence of labeling order of the countries. The figure clearly shows that US is the most affected country, it has highest confirmed and death cases. Except US and Russia, other countries are able to flatten the graphs after some level of outbreak.

\begin{figure}[htbp]
\centerline{\includegraphics[width=0.9\columnwidth]{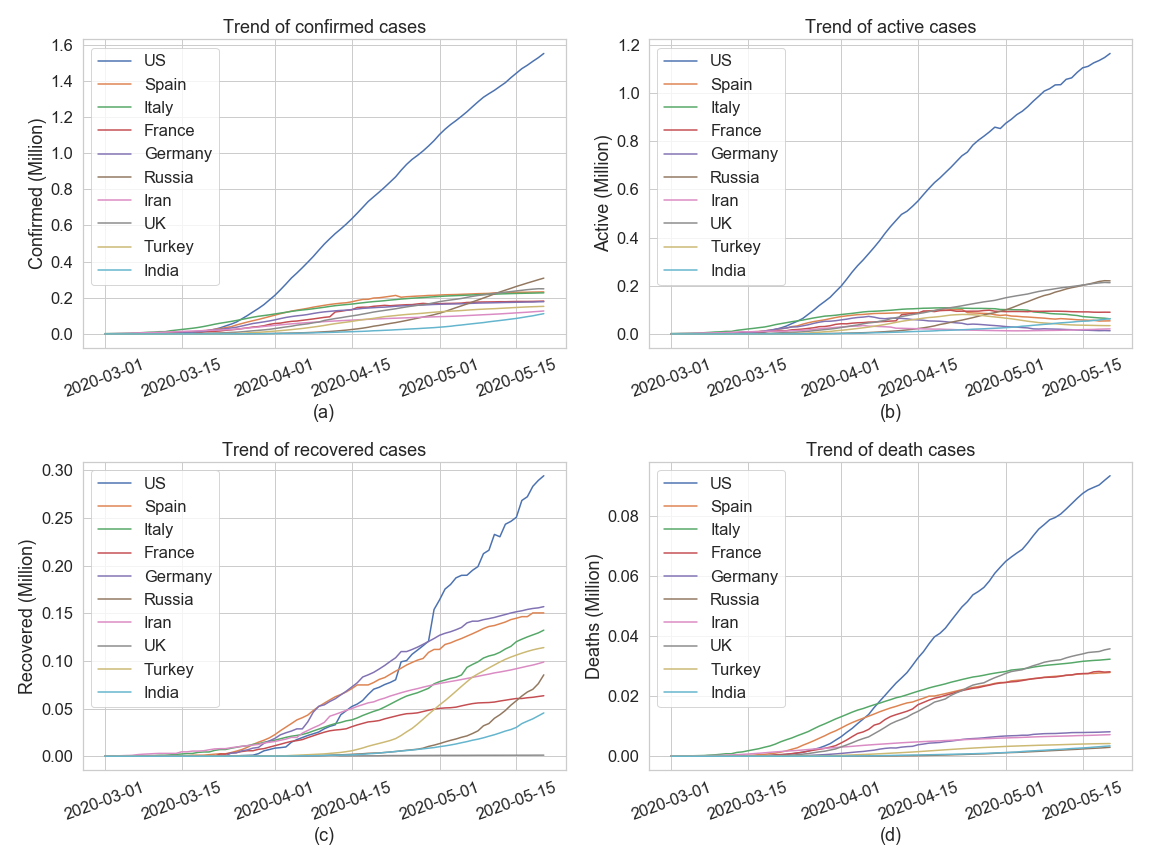}}
\caption{Trend of Confirmed, Recovered, Deaths, and Active cases of COVID-19 in adopted countries.}
\label{fig}
\end{figure}

Fig.  2 shows worldwide spread trend of coronavirus from 22 January, 2020 to 20 May, 2020. The trend explains that growth of the virus is almost exponential after mid-March 2020.

\begin{figure}[htbp]
\centerline{\includegraphics[width=0.6\columnwidth]{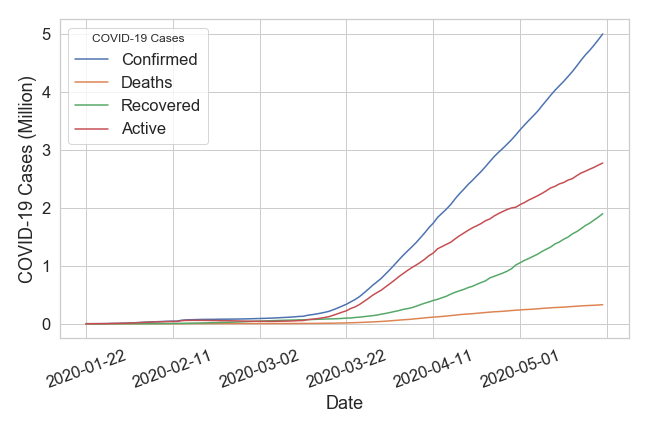}}
\caption{Trend of COVID-19 cases worldwide.}
\label{fig}
\end{figure}

Fig. 3 shows recovery rate of COVID-19 in the adopted countries and worldwide from 22 Jan 2020 to 20 May 2020. From the figure, we can say that highest recovery is done by Iran whereas Turkey shows exponential growth in the recovery cases, and other countries follows the recovery pattern similar to growth pattern of confirmed cases. Various studies \cite{sohrabi2020world, zhang2020predicting,haghani2020scientific} show that the disease recovers automatically after sometime but causes major health problem which can lead to death, if not taken care.

\begin{figure}[htbp]
\centerline{\includegraphics[width=0.9\columnwidth]{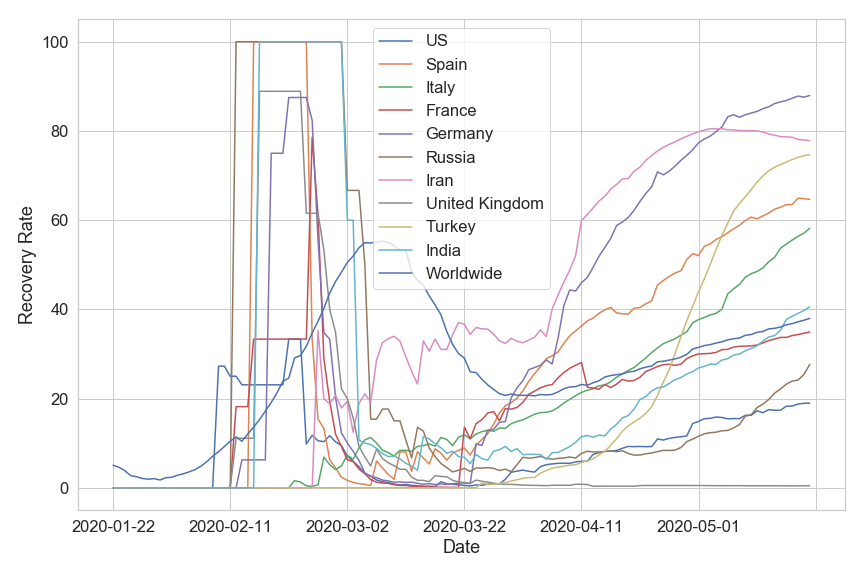}}
\caption{Recovery rate of COVID-19 cases.}
\label{fig}
\end{figure}

Fig. 4 shows fatality rate of COVID-19 patients worldwide and in adopted countries from 22 Jan 2020 to 20 May 2020. The figure shows that Iran faced highest death rate which was later over taken by US along with France. Spain has also shown significant death rate over the period of time. Other countries were able to control deaths to some extent using lockdowns or following social distancing etc. The virus has taken lives of many people worldwide.

\begin{figure}[htbp]
\centerline{\includegraphics[width=0.9\columnwidth]{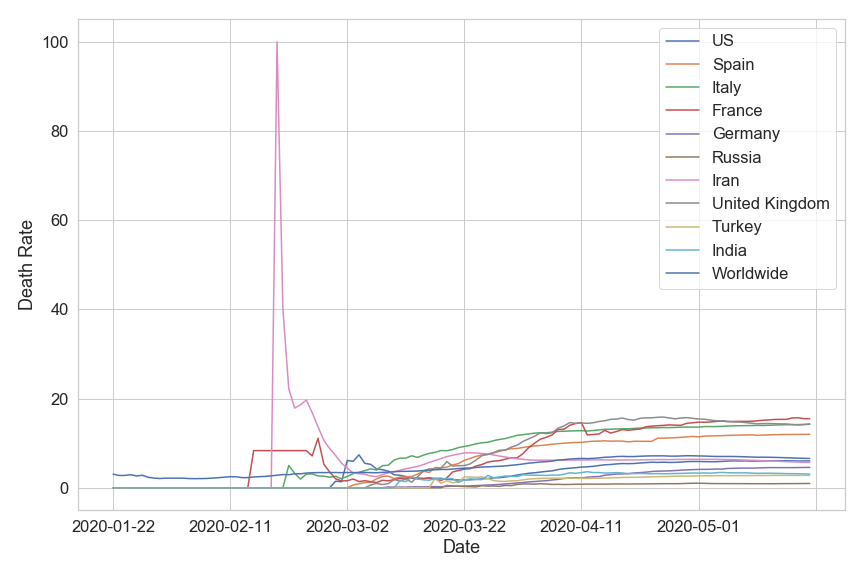}}
\caption{Fatality rate of COVID-19 cases.}
\label{fig}
\end{figure}

The historical data depicts that the COVID-19 badly affected the countries which do not impose lockdowns or do not followed social distancing. Some variations in virus spread rate, recovery rate, and death rate can be seen in different countries based on population density, available health system in a country, testing capability, and action taken to contain the outbreak.

\section{Time Series forecasting models}
Time series forecasting models are used to predict the futuristic outcomes based on historical information. We have adopted ARIMA and Facebook Prophet (FBProphet) model in our evaluative and forecasting study. An overview of the models is given in the following sections.

\subsection{Autoregressive Integrated Moving Average (ARIMA)}
ARIMA(p,d,q) \cite{wulff2017time} is composite of Autoregressive (AR) model, Moving Average (MA) model, and \textquotesingle I\textquotesingle{}   stands for integration; where p is order of autoregression, d is order of differencing, q is order of moving average.

The AR(p) model is defined as a linear process given as the following equation.

\begin{equation}
z_{t} =\alpha + \phi_{1} z_{t-1} + \phi_{2} z_{t-2} + ... + \phi_{p} z_{t-p} + w_{t}
\end{equation}

where $z_{t-1}, z_{t-2}, …z_{t-p}$ are the lags (past values); $\phi_{1}, \phi_{2}, ...\phi_{p}$ are lag coefficients which are estimated by the model; $w_{t}$ is the white noise, and $\alpha$ is defined as follows.

\begin{equation}
\alpha  = \left( 1- \sum_{i=1}^{p}\phi_{i} \right) \mu
\end{equation}

where $\mu$ is mean of the process. 

The MA(q) model is defined as the following equation.

\begin{equation}
z_{t} =\alpha + w_{t} + \theta_{1} w_{t-1} + \theta_{2} w_{t-2} + ... + \theta_{q} w_{t-q}
\end{equation}

Where $w_{t}, w_{t-1}, .. w_{t-q}$ are error terms of the model for the respective lags i.e. $z_{t}, z_{t-1}, ...z_{t-q}$ .

ARIMA is able to fit if the data is stationary i.e. data mean and standard deviation is constant. The differencing parameter d is the order of transformation to make dataset stationary. Second order differencing is shown in the following equation.

\begin{equation}
z_{t} = \left(Z_{t} - Z_{t-1}\right) - \left(Z_{t-1} - Z_{t-2}\right) = Z_{t} - 2Z_{t-1} + Z_{t-2}
\end{equation}

Finally the equation for the ARIMA(p,d,q) is defined as follows.

\begin{equation}
z_{t} =\alpha + \sum_{i=1}^{p}\phi_{i}z_{t-i}  +w_{t}+  \sum_{j=1}^{q}\theta_{j}w_{t-j}
\end{equation}

\subsection{Facebook Prophet}
Taylor \textit{et al.} \cite{taylor2018forecasting} proposed the Facebook Prophet (FBProphet) which uses several non-linear and linear methods as components with time as a regressor. Prophet is developed and released as open source software by data science team of Facebook. The model ignores the temporal dependence of the data, and training is framed just as curve-fitting exercise. So, irregular observations are also allowed in a dataset. The model offers various advantages like it can accommodate multiple period seasonality; it can accommodate custom and known holidays; it provides flexibility by offering two options for trend: 1. a piecewise linear model, 2. a saturating growth model; and the model fits very fast. The model includes one more term ‘holidays’ as the components of time series, so a time series can be defined by the following equation.

\begin{equation}
z_{t} = T_{t} + S_{t} + H_{t} + \epsilon_{t}
\end{equation}

where $T_{t}$ is trend, $S_{t}$ is seasonality, $H_{t}$ is holiday, and $\epsilon_{t}$ is error term. 

\section{Dataset and Forecasting framework}
This section describes about the dataset we have used to forecast COVID-19 cases of adopted countries, and worldwide. It also describes the modeling framework which we have followed.

\subsection{Modeling Dataset}
It is observed from the trends in section-III that rate of the reported COVID-19 cases in each country increases with time and flattens after sometime if large scale testing is performed,  lockdown is imposed, and containment is followed. For our study, we disaggregated the available day level data of the adopted 10 countries. We discarded the initial 5 days data for each country in our study. The reason to discarding the initial samples is that testing of the samples grows slowly in starting phase which does not depict the actual rate of the spread. The utilized samples detail is given in Table I. The end date of the collected samples is 20 May 2020. In this study, we consider 80 percentage samples for training and 20 percentage samples for testing the models for each country and worldwide data.

\begin{table}[htbp]
\caption{Total COVID-19 samples used for modeling till 20 May, 2020.}
\begin{center}
\renewcommand{\arraystretch}{1.2}
\scalebox{0.7}{
\begin{tabular}{|c|c|c|c|c|c|c|}
\hline 
\textbf{Region} & \textbf{Sample size (Days)}& \textbf{Start Date} & \textbf{Confirmed} & \textbf{Recovered} & \textbf{Deaths} & \textbf{Active}\\
\hline
Worldwide & 120	& 2020-01-22 & 4996472	& 1897466	& 328115 &	2770891 \\
\hline
US & 115	& 2020-01-27	& 1551853 &	294312 &	93439 &	1164102  \\
\hline
Spain & 105 &	2020-02-06 & 232555 & 150376 & 27888 & 54291  \\
\hline
Italy & 106	& 2020-02-05 & 227364 &	132282 & 32330 & 62752  \\
\hline
France & 113 & 2020-01-29 &	181700 & 63472 & 28135 & 90093  \\
\hline
Germany & 110 & 2020-02-01 & 178473 & 156966 & 8144 & 13363  \\
\hline
Russia &  106 &	2020-02-05 & 308705 & 85392 & 2972 & 220341 \\
\hline
Iran & 87 & 2020-02-24 & 126949 & 98808 & 7183 & 20958 \\
\hline
UK & 106 & 2020-02-05 &	249619 & 1116 & 35786 &	212717  \\
\hline
Turkey & 66	& 2020-03-16 & 152587 & 113987 & 4222 &	34378  \\
\hline
India &  107 & 2020-02-04 &	112028 & 45422 & 3434 &	63172 \\
\hline
\end{tabular}
\label{tab1}
}
\end{center}
\end{table}

\subsection{Forecasting Framework}
Fig. 5 describes about the adopted framework for prediction and analysis of the COVID-19 cases using ARIMA, and FBProphet models.
For the analysis, we have split the datasets of confirmed, active, recovered, and death cases into training and testing. We performed prediction after removing trends wherever applicable, and used statistical measures to evaluate the performance.

\begin{figure}[htbp]
\centerline{\includegraphics[width=0.5\columnwidth]{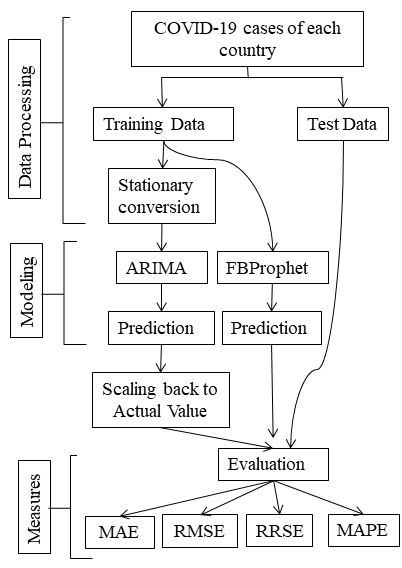}}
\caption{Framework to evaluate the forecasting models.}
\label{fig}
\end{figure}

\subsection{Performance Measures}
To evaluate the prediction models, we use the following statistical measures.

Mean Absolute Error (MAE):

\begin{equation}
MAE = \frac{1}{N}\sum_{k=1}^{N} \left| z_{k} - z\hat{}_{k} \right|
\end{equation}

Root Mean Square Error (RMSE):

\begin{equation}
RMSE = \sqrt{\frac{1}{N} \sum_{k=1}^{N} \left( z_{k} - z\hat{}_{k} \right)^{2} }
\end{equation}

Root Relative Squared Error (RRSE):

\begin{equation}
RRSE = \sqrt{\frac{\sum_{k=1}^{N} \left(z\hat{}_{k} - z_{k} \right)^{2}}{\sum_{k=1}^{N} \left( z\bar{} - z_{k} \right)^{2}} } \;\; where \; z\bar{} = \frac{1}{N} \sum_{k=1}^{N} z_{k}
\end{equation}

Mean Absolute Percentage Error (MAPE):

\begin{equation}
MAPE = \frac{100}{N} \sum_{k=1}^{N} \left| \frac{z_{k} - z\hat{}_{k}}{z_{k}} \right|
\end{equation}

where $z_{k}$ denotes actual value and $z\hat{}_{k}$ denotes predicted value for the $k_{th}$ instance.  $z\bar{}$ stands for the average value of  z, and N is the total number of testing samples.

\section{Results and discussion}
The adopted framework is implemented in Python 3.8, and we have used ARIMA and Prophet models from openly available packages \textit{statsmodels} and \textit{fbProphet} respectively. We have performed our experiments in Intel Core i5 processor clocked at 2.40 GHz, 8 GB RAM, and 4GB NVIDIA GTX-1650 GPU. In this section, we will discuss about forecasting accuracy of adopted models for active, recovered, deaths, and confirmed cases.

\subsection{Forecasting of active cases}
Active cases are the number of infected people who are under medical supervision. Active cases are derived as shown in the following equation. 

\begin{equation}
Active = Confirmed - Recovered - Deaths
\end{equation}

We use ARIMA and FBProphet models to predict the future cases. ARIMA can be used for prediction if data is stationary. It is clear from the trends in section-III that data of active cases is not in stationary form. So, we have applied techniques to convert the data into stationary form for ARIMA evaluation. We have applied square root scaling and one lag differencing to convert the data into stationary form. We have performed dicky-fuller test to check stationarity of the data. We also have used PACF and ACF plots to identify appropriate values of q and p order of ARIMA. The FBProphet is applied directly on actual data. Forecasting accuracy results for active cases of 10 adopted countries and worldwide are shown in Table II. We have mentioned order of ARIMA along with the accuracy results in the tables. The mentioned ARIMA order performed better to fit the model accurately. Best MAPE scores are 0.586 and 1.481 for US and UK data by ARIMA and FBProphet respectively. From the results, we can clearly say that ARIMA has far better performance as compared to FBProphet model with respect to all types of error measures i.e. MAE, RMSE, RRSE, and MAPE.

\begin{table}[htbp]
\caption{Performance results of the models for COVID-19 active cases in adopted countries.}
\begin{center}
\renewcommand{\arraystretch}{1.2}
\scalebox{0.8}{
\begin{tabular}{|c|c|c|c|c|c|}
\hline
\textbf{Region} & \textbf{Model} & \textbf{MAE} & \textbf{RMSE} & \textbf{RRSE} & \textbf{MAPE} \\
\hline
\multirow{2}{*}{Worldwide} & ARIMA(9,1,2) & 19141.89 & 21377.14	& 0.086	& 0.816 \\
\cline{2-6}
 & FBProphet & 168452.05 & 182230.63 & 0.706 & 6.943  \\
\hline
\multirow{2}{*}{US} & ARIMA(10,1,3)	& 5732.16 &	8050.31 & 0.079	& 0.586 \\
\cline{2-6}
 & FBProphet & 95766.22	& 108424.76	& 1.07	& 9.12  \\
\hline
\multirow{2}{*}{Spain} & ARIMA(8,1,4)	& 2191.68 & 2603.02 & 0.346 & 3.293 \\
\cline{2-6}
 & FBProphet & 67132.86	& 69748.42 & 9.274 & 109.40  \\
\hline
\multirow{2}{*}{Italy} & ARIMA(9,1,3) &	3197.25	& 4266.60 &	0.320 & 3.411 \\
\cline{2-6}
 & FBProphet & 26934.34	& 30963.76	& 2.325 & 35.55  \\
\hline
\multirow{2}{*}{France} & ARIMA(5,1,4)	& 10974.15	& 11489.85	& 6.166	& 11.75 \\
\cline{2-6}
 & FBProphet & 44596.16	& 48195.48	& 25.864 &	48.340  \\
\hline
\multirow{2}{*}{Germany} & ARIMA(11,1,4) &	2114.09	& 2597.193 & 0.407 & 9.052  \\
\cline{2-6}
 & FBProphet & 50902.42	& 52259.90 &	8.197 &	277.26  \\
\hline
\multirow{2}{*}{Russia} & ARIMA(10,1,2)	& 6456.26	& 6786.96 &	0.158 &	4.238  \\
\cline{2-6}
 & FBProphet & 36430.36	& 40232.57 & 0.936 & 20.748  \\
\hline
\multirow{2}{*}{Iran} & ARIMA(4,1,2) & 328.28 & 379.79 & 0.147 & 2.202 \\
\cline{2-6}
 & FBProphet & 12856.19	& 12902.11 & 5.009 & 82.503  \\
\hline
\multirow{2}{*}{UK} & ARIMA(4,1,2)	& 8090.84 &	8637.25	& 0.375	& 4.66 \\
\cline{2-6}
 & FBProphet &  2954.65	& 4649.43 &	0.202 &	1.481  \\
\hline
\multirow{2}{*}{Turkey} & ARIMA(8,1,2) & 3631.37 &	3655.74 & 0.884	& 9.485 \\
\cline{2-6}
 & FBProphet & 59801.55	& 60725.11 & 14.678	& 158.59  \\
\hline
\multirow{2}{*}{India} & ARIMA(11,1,5)	& 7007.09 &	7330.06	& 0.61 & 16.74 \\
\cline{2-6}
 & FBProphet & 10245.17	& 12085.37 & 1.005 & 21.429  \\
\hline

\end{tabular}
\label{tab1}
}
\end{center}
\end{table}

\begin{table}[htbp]
\caption{Performance results of the models for COVID-19 recovered cases in Adopted countries.}
\begin{center}
\renewcommand{\arraystretch}{1.2}
\scalebox{0.8}{
\begin{tabular}{|c|c|c|c|c|c|}
\hline
\textbf{Region} & \textbf{Model} & \textbf{MAE} & \textbf{RMSE} & \textbf{RRSE} & \textbf{MAPE} \\
\hline
\multirow{2}{*}{Worldwide} & ARIMA(9,1,2) &	34932.99 &	36992.53 &	0.128 &	2.523  \\
\cline{2-6}
 & FBProphet & 185584.71 &	214741.61 &	0.712 &	12.49  \\
\hline
\multirow{2}{*}{US} & ARIMA(5,1,2) & 31899.89 &	33109.68 &	0.667 &	15.635 \\
\cline{2-6}
 & FBProphet &  53970.19 & 57816.45 & 1.165 & 24.174  \\
\hline
\multirow{2}{*}{Spain} & ARIMA(8,1,4) & 9683.45 & 9774.06 &	0.786 &	7.361 \\
\cline{2-6}
 & FBProphet &  3021.53	& 3766.69 &	0.303 &	2.22  \\
\hline
\multirow{2}{*}{Italy} & ARIMA(9,1,3) &	12910.06 &	13078.23 &	0.693 &	12.78  \\
\cline{2-6}
 & FBProphet &  8721.87	& 10057.88 & 0.533 & 7.881  \\
\hline
\multirow{2}{*}{France} & ARIMA(3,1,1) & 5780.87 &	5853.29	& 1.21	& 10.574 \\
\cline{2-6}
 & FBProphet &  7323.90	& 8362.88 &	1.729 &	12.613  \\
\hline
\multirow{2}{*}{Germany} & ARIMA(5,1,3) & 13702.61 & 13901.04 &	1.287 &	9.808 \\
\cline{2-6}
 & FBProphet &  25017.26 &	28763.20 &	2.664 &	16.969  \\
\hline
\multirow{2}{*}{Russia} & ARIMA(4,1,0) & 2376.69 &	3212.50	& 0.141	& 5.103 \\
\cline{2-6}
 & FBProphet & 26988.80	& 33858.56 & 1.484 & 60.964  \\
\hline
\multirow{2}{*}{Iran} & ARIMA(1,1,1) &	4213.14 & 4496.75 &	0.736 &	4.933 \\
\cline{2-6}
 & FBProphet & 5638.72 & 6037.87 &	0.988 &	6.267  \\
\hline
\multirow{2}{*}{UK} & ARIMA(4,1,2) & 78.19 & 91.12 & 1.177 & 8.311  \\
\cline{2-6}
 & FBProphet &  69.11 &	79.44 &	1.026 &	7.326  \\
\hline
\multirow{2}{*}{Turkey} & ARIMA(8,1,2) & 4242.09 & 4333.57 & 0.44 &	4.321 \\
\cline{2-6}
 & FBProphet &  45986.27 &	46211.42 &	4.688 &	45.536  \\
\hline
\multirow{2}{*}{India} & ARIMA(2,1,0) &	721.17 & 1066.65 &	0.096 &	2.911 \\
\cline{2-6}
 & FBProphet &  11395.90 &	14381.55 &	1.295 &	42.882  \\
\hline

\end{tabular}
\label{tab1}
}
\end{center}
\end{table}

\begin{table}[htbp]
\caption{Performance results of the models for COVID-19 fatality cases in adopted countries}
\begin{center}
\renewcommand{\arraystretch}{1.2}
\scalebox{0.8}{
\begin{tabular}{|c|c|c|c|c|c|}
\hline
\textbf{Region} & \textbf{Model} & \textbf{MAE} & \textbf{RMSE} & \textbf{RRSE} & \textbf{MAPE} \\
\hline
\multirow{2}{*}{Worldwide} & ARIMA(9,1,2) &	661.98 & 821.20 & 0.026	& 0.257  \\
\cline{2-6}
 & FBProphet &  21666.12 &	24874.45 &	0.735 &	7.465 \\
\hline
\multirow{2}{*}{US} & ARIMA(2,1,0) & 1924.08 &	1988.71	& 0.19 & 2.571 \\
\cline{2-6}
 & FBProphet &  4799.16 &	5856.54 &	0.56 &	5.751  \\
\hline
\multirow{2}{*}{Spain} & ARIMA(2,1,0) &	940.85 & 953.08 & 0.907 & 3.577 \\
\cline{2-6}
 & FBProphet & 2573.67 & 2961.76 &	2.818 &	9.525  \\
\hline
\multirow{2}{*}{Italy} & ARIMA(2,1,0) &	1240.10 & 1254.32 &	0.892 &	4.128 \\
\cline{2-6}
 & FBProphet &  3008.94 &	3433.46 & 2.443	& 9.703 \\
\hline
\multirow{2}{*}{France} & ARIMA(3,1,1) & 1335.79 &	1355.02 & 0.983 & 5.139 \\
\cline{2-6}
 & FBProphet & 6545.93 & 7270.98 &	5.274 &	24.382  \\
\hline
\multirow{2}{*}{Germany} & ARIMA(1,1,0) & 318.04 &	341.57 & 0.668 & 4.382 \\
\cline{2-6}
 & FBProphet &  1446.64 &	1668.89 & 3.262	& 18.761  \\
\hline
\multirow{2}{*}{Russia} & ARIMA(2,1,0) & 43.31 & 48.98 & 0.082  & 2.252 \\
\cline{2-6}
 & FBProphet &  628.39 & 709.50 & 1.184 & 30.597  \\
\hline
\multirow{2}{*}{Iran} & ARIMA(1,1,1) &	836.66 & 836.86	& 2.929 & 12.487  \\
\cline{2-6}
 & FBProphet & 257.44 &	291.70 & 1.021 & 3.759  \\
\hline
\multirow{2}{*}{UK} & ARIMA(2,1,0) & 959.53	& 984.02 &	0.343 &	3.119  \\
\cline{2-6}
 & FBProphet &  4171.84 &	4867.84 & 1.699 & 12.639  \\
\hline
\multirow{2}{*}{Turkey} & ARIMA(8,1,2) & 113.54 & 117.61 & 0.619 &	2.909  \\
\cline{2-6}
 & FBProphet & 280.83 &	312.96 & 1.647 & 6.945  \\
\hline
\multirow{2}{*}{India} & ARIMA(2,1,0) &	48.94 &	60.75 &	0.085 &	2.704 \\
\cline{2-6}
 & FBProphet &  771.35 & 897.58	& 1.26 & 31.822 \\
\hline

\end{tabular}
\label{tab1}
}
\end{center}
\end{table}

\subsection{Forecasting of recovered cases}
To predict and analyse the recovery rate of the disease, we have performed evaluative study of the adopted models using recovery data of the 10 adopted countries, and worldwide. We can see from the trends in section III that recovery data is also non-stationary. So, we have performed stationarity techniques similar to as discussed in section-VI(A) to evaluate the ARIMA model. We have applied FBProphet directly on actual data to fit the model and generated the forecasting results. Table III shows accuracy results of the models for the recovered cases. Best MAE results are 78.19 and 69.11 for UK data by ARIMA and FBProphet respectively. Results show that ARIMA prediction almost matches the actual values whereas FBProphet did not perform as well. We can see that maximum MAPE value of ARIMA is 15.6 and minimum value is 2.5 which are very much acceptable to generate forecasting results, whereas maximum and minimum MAPE for FBProphet are 31.822 and 3.759 respectively.

\subsection{Forecasting of death cases}
Coronavirus has taken many lives. So, it is necessary to analyse the fatality rate of the virus, and forecasting to highlight future cases which can guide governments to act in advance. In this section, we have evaluated the forecasting models for death cases of the adopted countries, and worldwide. We have converted the non-stationary fatality data into stationary form to fit the ARIMA model similar to as discussed in section-VI(A). FBProphet model is applied on the actual data to forecast the prediction results. Table IV shows prediction accuracy of the models for the fatality cases. We can see that prediction errors of ARIMA are very less whereas FBProphet prediction have high error factor in the results. The results suggest that ARIMA can be used for actual forecasting of the cases to plan the services accordingly.

\begin{figure}[htbp]
    \centering
  \begin{subfigure}{8cm}
    \centerline{\includegraphics[width=0.8\columnwidth]{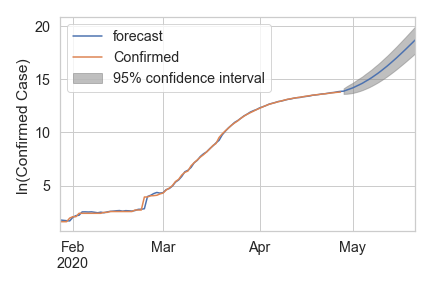}}
    \caption{ARIMA Forecasting for US confirmed cases}
  \end{subfigure}
  \begin{subfigure}{8cm}
    \centerline{\includegraphics[width=0.8\columnwidth]{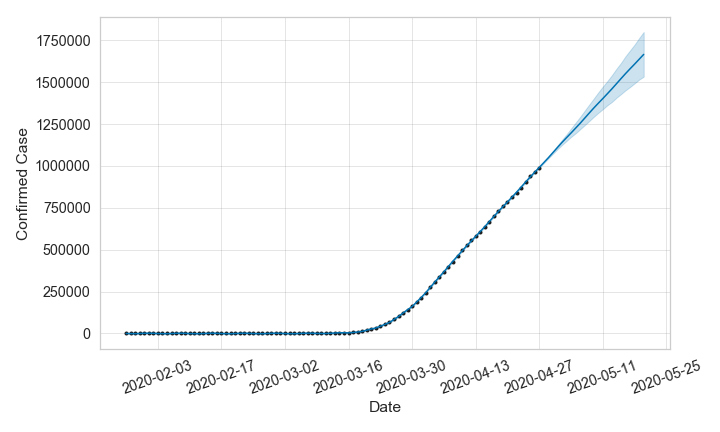}}
    \caption{FBProphet Forecasting for US confirmed cases}
  \end{subfigure}
  \caption{Actual and predicted values plots of ARIMA and FBProphet for covid-19 confirmed cases of US.}
  \label{fig}
  \end{figure}
  
\subsection{Forecasting confirmed cases}
In this section, we have highlighted the fitted accuracy of the models using confirmed cases. For this analysis, we have chosen only two countries; US and India. The results of US and India by both the models ARIMA and FBProphet are shown in Fig. 6 and Fig. 7 respectively. We have shown training and testing data split using vertical line in the figures. Forecasted and actual data are plotted together to visualize the fitting accuracy of the models. We can see that FBProphet model is able to fit well in case of US data as shown in Fig. 6 whereas ARIMA is able to perform well in case of India data as shown in Fig. 7. FBProphet adopts successively progression, and avoid outliers during modeling and forecasting. The results also depicts that FBProphet can fit well in case of less data whereas ARIMA requires sufficient data to model and predict the results.

\begin{figure}[htbp]
    \centering
  \begin{subfigure}{8cm}
    \centerline{\includegraphics[width=0.8\columnwidth]{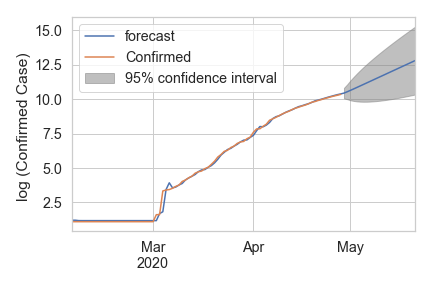}}
    \caption{ARIMA Forecasting for India confirmed cases}
  \end{subfigure}
  \begin{subfigure}{8cm}
    \centerline{\includegraphics[width=0.8\columnwidth]{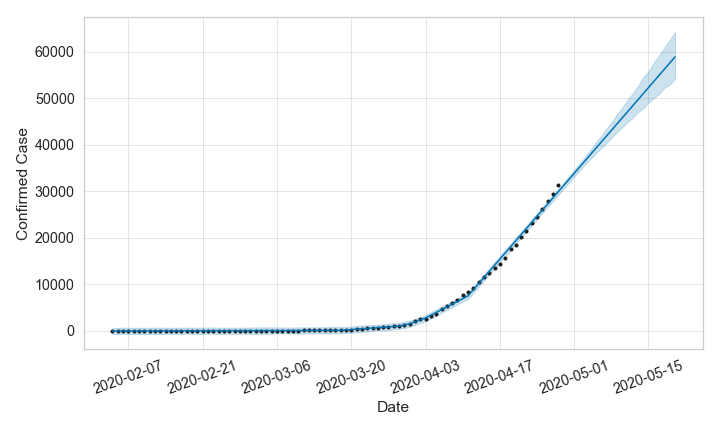}}
    \caption{FBProphet Forecasting for India confirmed cases}
  \end{subfigure}
  \caption{Actual and predicted values plots of ARIMA and FBProphet for covid-19 confirmed cases of India.}
  \label{fig}
  \end{figure}

\section{Conclusion and Future work}
WHO has declared COVID-19 as pandemic because it has infected most of the countries, and it is a major threat to human race. In this paper, we have done analysis and prediction study of the disease using widely accepted forecasting models; ARIMA and FBProphet. We have collected COVID-19 data of 10 highly affected countries US, Spain, Italy, France, Germany, Russia, Iran, UK, Turkey, India, and worldwide latest by May 20, 2020. For the most of the countries data, ARIMA has better performed compared to Prophet on scale of MAE, RMSE, RRSE, and MAPE error matrices. The trend analysis shows rapid growth in the infected cases, and prediction study shows great rise in the expected active, recovered, and death cases worldwide. However, lockdowns and containment policies may affect the prediction results. The adopted models have performed well but it limits our study to the effectiveness of the models, which can be further improved using ensemble of multiple prediction models. The obtained forecasting results further can be improved by taking various variables into account like population density, weather, health system, patient history etc. using deep learning techniques, and artificial intelligence.

\bibliographystyle{plain}
\bibliography{covid-19-forecasting}

\end{document}